\documentclass{aastex}          %% The manuscript based on AASTeX v5.x
\usepackage{spr-astr-addons}    %% mimicing ASTR journal style
\usepackage{multirow,epstopdf}             %for PS/EPS graphics inclusion, new
\usepackage[Symbol]{upgreek}

\begin{document}
%% Article title
%
\title{Fringes' Impacts to Astrometry and Photometry of Stars}

%% Running heads
\shorttitle{<Short article title>}
\shortauthors{<Autors et al.>}

%% Author and Affilations
\author{Z. J. ~Zheng\altaffilmark{1,3}}
\and
\author{Q. Y. ~Peng\altaffilmark{2,3}}
%\affil{}
%\email{} %% non-output

%% Alternate Affilations
\altaffiltext{1}{Computer Center, Guangdong University of
Petrochemical Technology, Maoming 525000, China;}
\altaffiltext{2}{Dept. of Computer Science, Jinan University, Guangzhou 510632, China;}
\altaffiltext{3}{Sino-France Joint Laboratory for Astrometry, Dynamics and Space Science,
Jinan University, Guangzhou 510632, China}

% Abstract
\begin{abstract}
 Fringes often appear in a CCD frame, especially when a thin CCD chip and a $R$ or $I$ filter is used. 88 CCD frames of the two open clusters NGC 2324 and NGC 1664 with a Johnson $I$ filter taken from the 2.4-m telescope at Yunnan Observatory are used to study the fringes' impacts to the astrometry and photometry of stars. A novel technique proposed by Snodgrass \& Carry is applied to remove the fringes in each CCD frame. And an appraisal of this technique is performed to estimate fringes' effects on astrometry and photometry of stars. Our results show that the astrometric and photometric precisions of stars can be improved effectively after the removal of fringes, especially for faint stars.
\end{abstract}

% Keywords

\section{Introduction}           %% first-level sections will be auto-capitalized
\label{sect:intro}

Large-aperture optical telescopes have a potential for deep-sky exploration with their powerful capability to collect the flux of light. However, the faint stars are subject to some systematic errors, such as bad pixels or columns, quantum efficiency variations, fringes, etc. Though some of these systematic errors can be eliminated by using standard data-processing techniques (bias subtraction, flat-fielding, bad pixel masking, etc), more special care must be taken for fringes' removal, since the fringes are related with the CCD itself and a light wavelength. When a wavelength beyond 600$\sim$700 nm, the absorption efficiency in the CCD silicon gradually decreases with the increase of wavelength. This would lead to fringes, resulting from multiple reflections and interference between CCD surfaces. The variation of CCD surface's thickness would lead to same variation in fringe's amplitude, phase and quantum efficiency, as a function of pixel position on the CCD frame \citep{Wong10}. Besides, the observed fringe pattern also depends on the wavelength range of the light. Monochromatic illumination with a proper wavelength produces strong fringe pattern while broad-band illumination produces weaker fringes as the range of wavelengths washes out the appearance of fringes \citep{Wong10}. The night-sky emission lines are strong particularly in the red part of the optical spectrum (e.g. the I-band) and primarily drive the fringe pattern in ground-based observations.

As such, the fringe pattern changes little over time with the same filter. It can be derived by median filtering of dithering exposures' stack in quantity \citep{Gullixson92} or neon lamp flat-fielding \citep{Howell12}. Nonetheless, fringes' amplitudes vary from frame to frame, most likely depending on exposure times, air-masses and weather conditions, etc. Snodgrass \& Carry (\citeyear{Snodgrass13}) presented a simple but effective way, through referring a series of pixel pairs, to remove the fringes in original frames automatically.

Fringes' effects were studied systematically based on some flat field frames, which were used by the WFC3 calibration pipeline \citep{Wong10}. Within a series of flat field data in different filters, the F953N flat fields showed strongest fringes \citep[Fig.19]{Wong10}. Since the lack of a neon lamp, we prefer to study fringes' impacts based on actual science data rather than laboratory experiments.

In this paper, we carry out a study of fringes' effects to some practical observations for their astrometry and photometry based on the CCD frames taken from the 2.4-m telescope at Yunnan Observatory. The geometric distortion (called GD hereafter) correction is also applied for their high-precision astrometry \citep{Peng12}.

The contents of this paper are arranged as follows. In Section 2, the observations are described. In Section 3, we give details on the data reduction, mainly concentrating on defringing procedure and GD solution. The results for astrometry and photometry are discussed in Section 4 and Section 5 respectively. Finally, the conclusions are drawn in Section 6.

\section{Observations}
\label{sect:Obs}

Our observations were obtained from the 2.4-m telescope with an E2V CCD42-90 chip at Yunnan Observatory on January 3, 2011. Two open clusters NGC 2324 and NGC 1664 were observed by using a dithering scheme. There were 44 exposures for each of the open clusters in a Johnson $I$ filter. The typical seeing of the observations is about $1{\arcsec}\!.5$ (FWHM). The brightest star on a frame was just saturated and the exposure time were 30 seconds for NGC 2324 and 19 seconds for NGC 1664. And a clip process was applied for the raw CCD frames to avoid the ineffective boundary, leaving an area of 1900${\times}$1900 pixels. Specifications of the 2.4-m telescope and its CCD chip are listed in Table~\ref{Tab1}.

 \begin{table}
 \caption{Specifications of the 2.4-m Telescope and CCD
Detector} %% no full stop at the end of caption
 \label{Tab1}
 \begin{tabular}{rr}
 \tableline  %% rule at top
% \tablenotemark{}
Approximate focal length             &1920 cm\\
F-Ratio                           &8\\
Diameter of primary mirror        &240 cm\\
CCD field of view (effective)           &$\approx$9${\arcmin}$${\times}$9${\arcmin}$\\
Size of CCD array (effective)         &1900${\times}$1900\\
Size of pixel               &13.5${\times}$13.5~$\upmu$m$^2$ \\
Approximate scale factor  &0.286 arcsec~pixel$^{-1}$\\
 \tableline %% rule at bottom
 \end{tabular}
 \end{table}

\section{Data reduction}
\label{sect:reduction}
Before the further reduction, a series of calibrations are done: (1) bias subtraction, flat-fielding and removal of cosmic rays; (2) fringes' removal; (3) derive the GD pattern and correct the pixel positions of stars by using it. The details of defringing and GD solution are given as the subsections.

\subsection{REMOVAL OF FRINGES}
In order to derive a fringe pattern, we stack a series of frames which is taken by a dithering scheme. A median filter is applied at each pixel position of the stack and finally a fringe pattern is composed of every median value \citep{Gullixson92}. However, only 44 frames of NGC 2324 are selected to derive the fringe pattern because their signal-to-noise (S/N) is higher than the CCD frames of NGC 1664.

We follow the procedure first developed by Snodgrass and Carry (\citeyear{Snodgrass13}) to derive fringes' amplitude of an original frame and the positions of pixel pairs are chosen to avoid the errors resulted from the flux of stars. To be specific, a series of pixel pairs are set between fringes' bright area and dark area on the original frame. Once again, the pixel pairs are set at the same position of the fringe pattern (the red lines on the left panel of Figure~\ref{Fig1}). For the $i^\mathrm{th}$ pixel pair, the flux difference between bright and dark areas on the original frame (noted as $O$) and on the fringe pattern (noted as $F$) is calculated as follows:
{\setlength\abovedisplayskip{1pt}
\setlength\belowdisplayskip{1pt}
\begin{subequations}
\begin{align}
  \delta{O}_i&=O^i_{bright}-O^i_{dark}\\
  \delta{F}_i&=F^i_{bright}-F^i_{dark}
\end{align}
\label{eq:1}
\end{subequations}
{\setlength\abovedisplayskip{1pt}
\setlength\belowdisplayskip{1pt}
We usually set 30$\sim$40 pixel pairs on an original frame for its fringes' removal. After 3$\sigma$-clip, the list of $\delta{O}_i/\delta{F}_i$ ($i=1,2,\cdots,N$) for an original frame is shown on the right panel of Figure~\ref{Fig1}. We take the median of the $\delta{O}_i/\delta{F}_i$ as the fringe scaling factor to remove some large discrepancies with a high probability of being outliers. An example of removing fringes is shown in Figure~\ref{Fig2}.

\begin{figure*}[tb]
\centering
\includegraphics[width=0.36\textwidth]{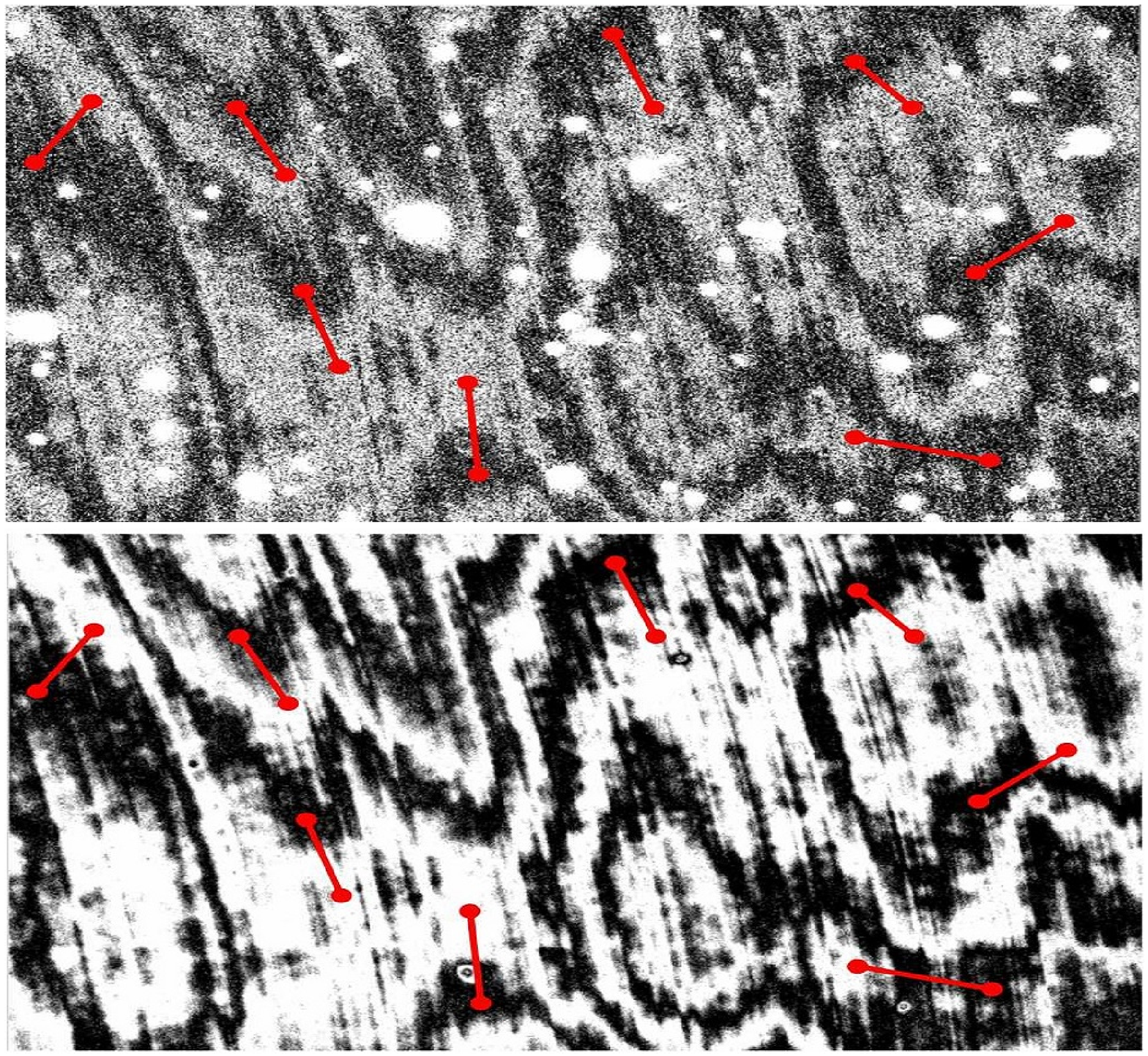}
\includegraphics[width=0.44\textwidth]{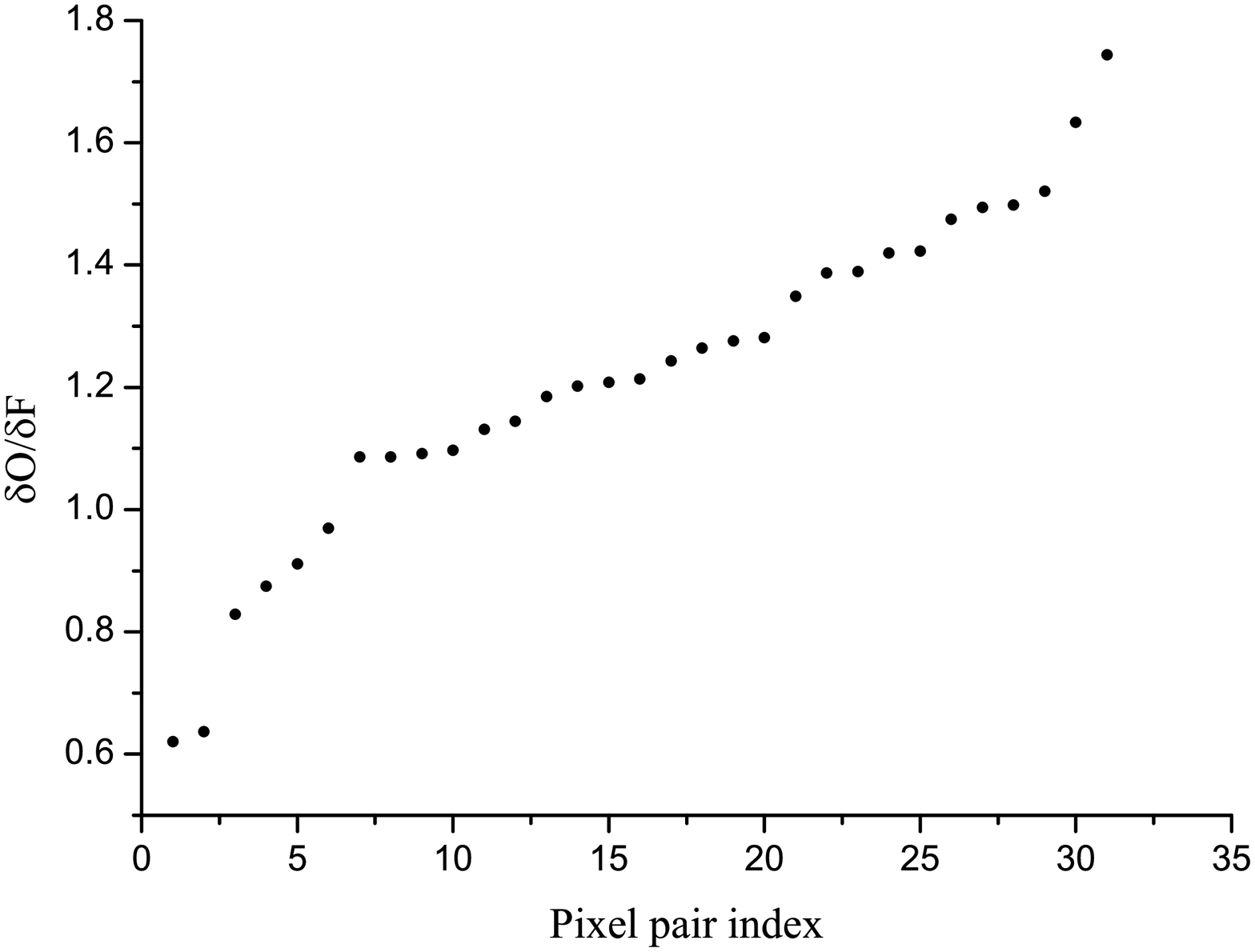}
\caption{Left: set the pixel pairs on the frames; Right: distribution of $\delta{O}_i/\delta{F}_i$ for an original frame.}
\label{Fig1}
\end{figure*}
\begin{figure*}
\includegraphics[width=0.33\textwidth]{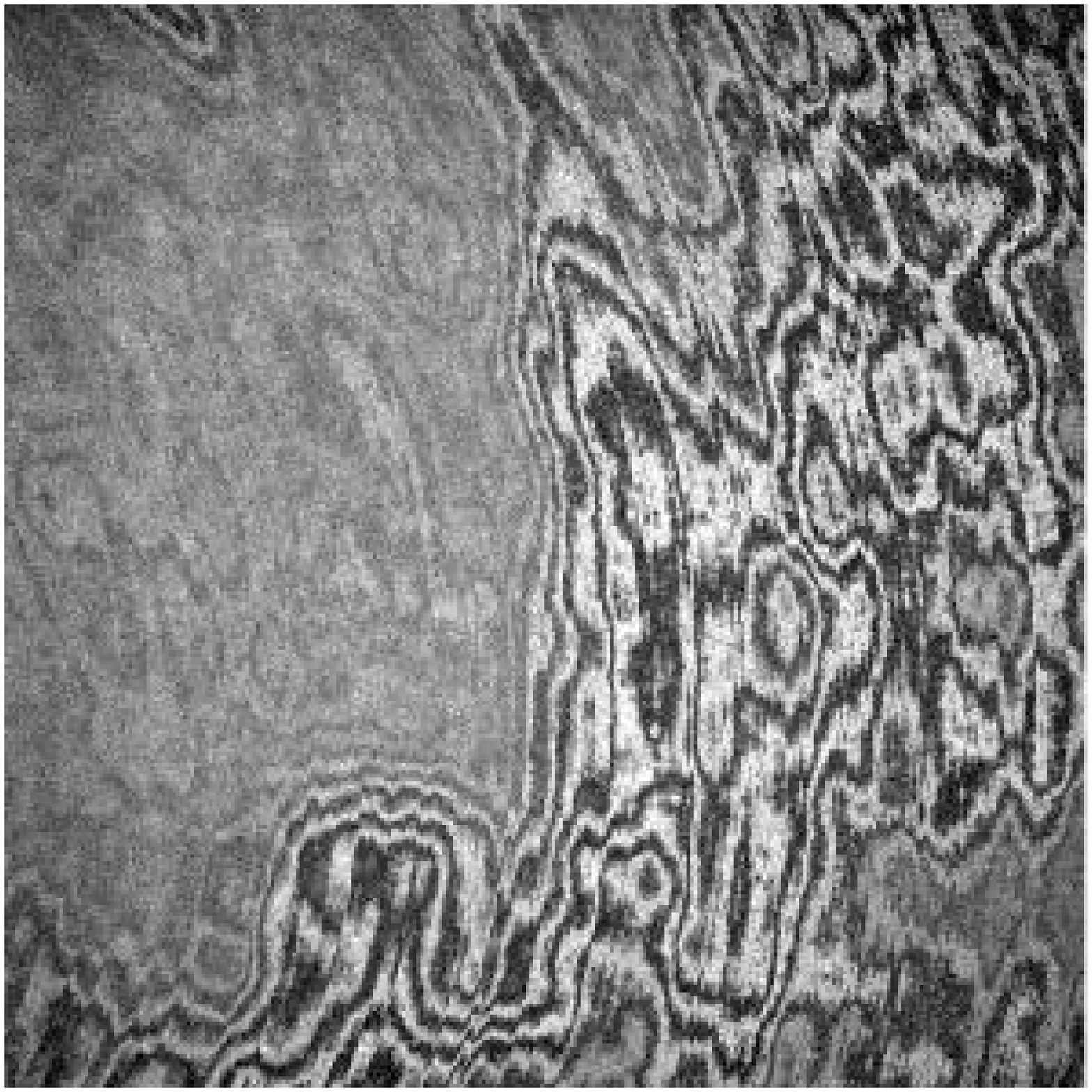}
\includegraphics[width=0.33\textwidth]{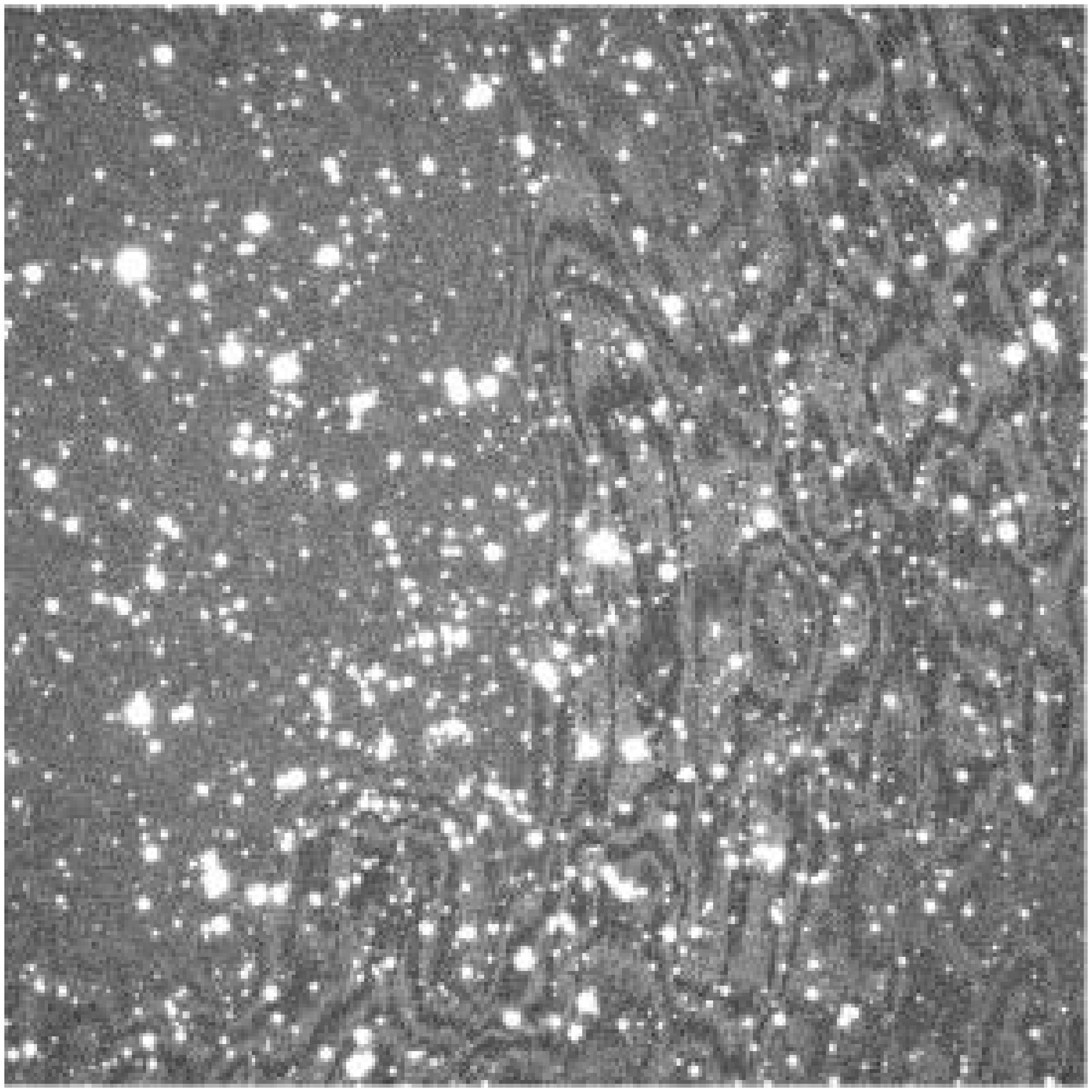}
\includegraphics[width=0.33\textwidth]{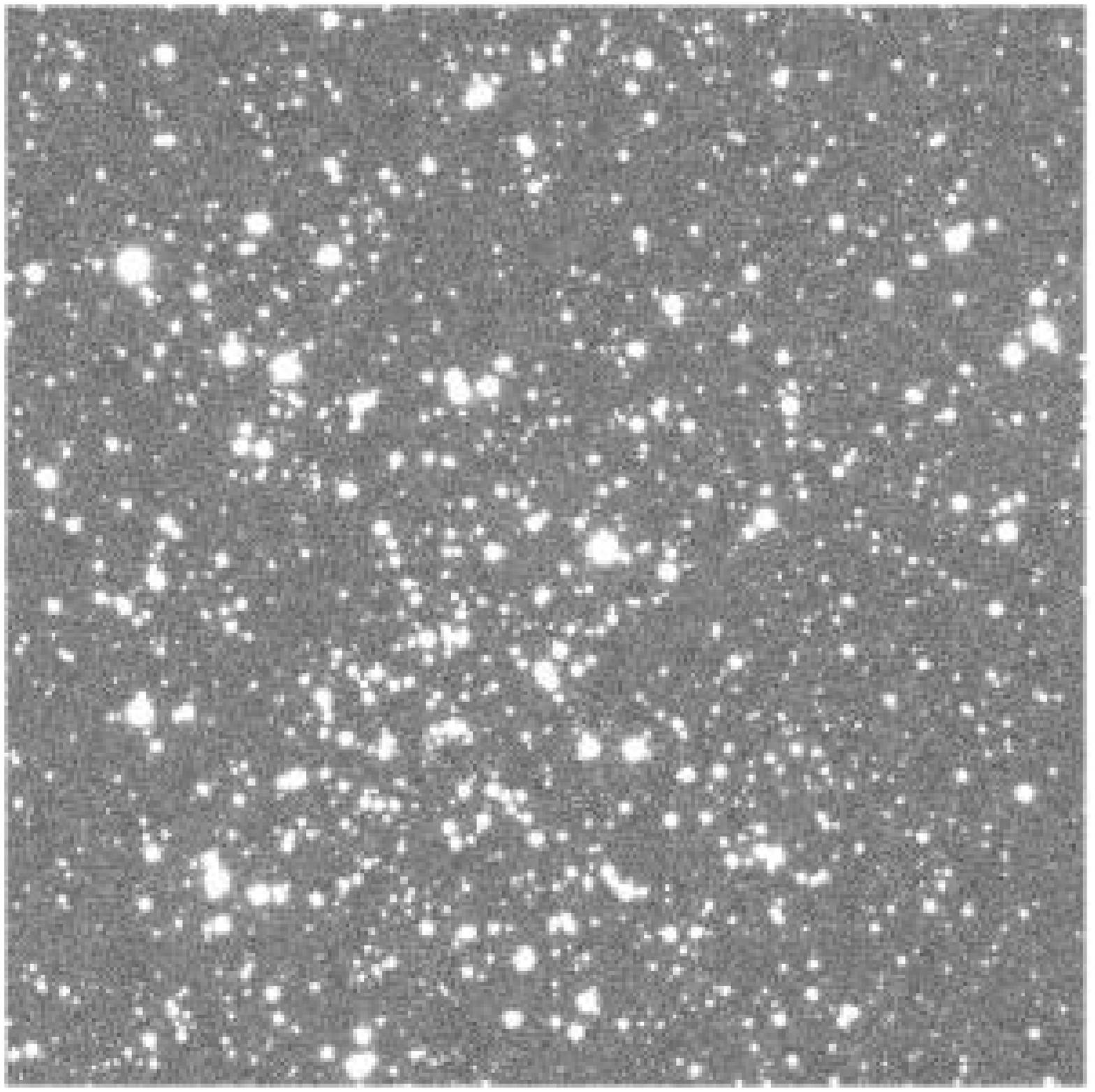}
\caption{Left: the fringe pattern; Middle: an original frame; Right: a defringed frame.}
\label{Fig2}
\end{figure*}

\subsection{SOLUTION OF GD}

Since the optical system of a telescope is not a perfect pin-hole model, there are more or less GD effects inevitably. For the 2.4-m telescope at Yunnan Observatory, whose effective field-of-view is only 9${\arcmin}$${\times}$9${\arcmin}$, the maximum GD would reach up to 1 pixel \citep{Zhang12}. In order to preserve a good astrometry precision, we solve the pattern of GD depending on the defringed frames as follows.
First, a two-dimensional Gaussian fitting and aperture photometry are applied to measure a star's pixel position and instrumental magnitude (the zero point is 25$^\mathrm{th}$ magnitude). Then we identify it according to the PPMXL catalog \citep{Roeser10} which can be downloaded from the VizieR database. About 800 reference stars are matched on each frame. The theoretical positions of reference stars from the catalog are transformed to pixel positions of the frames, through a four-parameter model after considering all astrometry effect (topocentric apparent position, atmospheric refraction, etc). The difference between the directly measured positions from frames and the indirectly computed ones from the catalog (observed minus calculated; O $-$ C), can be resolved into three sources: GD, catalog error and measured error. As the same star in dithering exposures falls in different pixel positions on the CCD chip, the GD in a specific pixel positions can be derived by canceling out the catalog errors and compressing the measured errors \citep{Peng12}.

The field of view of the CCD chip is divided into 19$\times$19 cells (100$\times$100 pixels per cell) and the GD in each cell is solved as follows. As mentioned above, if the same star is located at different positions in many CCD frames, its GD in a specific cell can be solved from the mean of differences of (O $-$ C) residuals. The remaining GD in a cell can be estimated by the mean of GDs, which are derived from stars falling in the cell. In each iteration, the observed pixel positions are updated through a bilinear interpolation from the newly-derived GD pattern and the GD pattern is re-calculated, until the corrections of GD are under a given threshold, such as 0.01 pixel. Left panel of Figure~\ref{Fig3} shows the final GD pattern. We also solve the GD pattern from the original frames as shown in the middle panel, which is similar to the left. Finally, we make the subtraction of the two GD patterns (shown in the right panel) to check the effect of defringing for GD solution. There are negligible differences in most cells for the two GD patterns, except in the lower right corner of the field of view, which can reach up to 0.1 pixel ($\approx$0.029 arcsec). Although fringes would cause an uneven sky background, we find that, the removal of fringes has only negligible effect to GD solution since only faint stars are affected by the fringes.
\begin{figure*}
\includegraphics[width=0.33\textwidth]{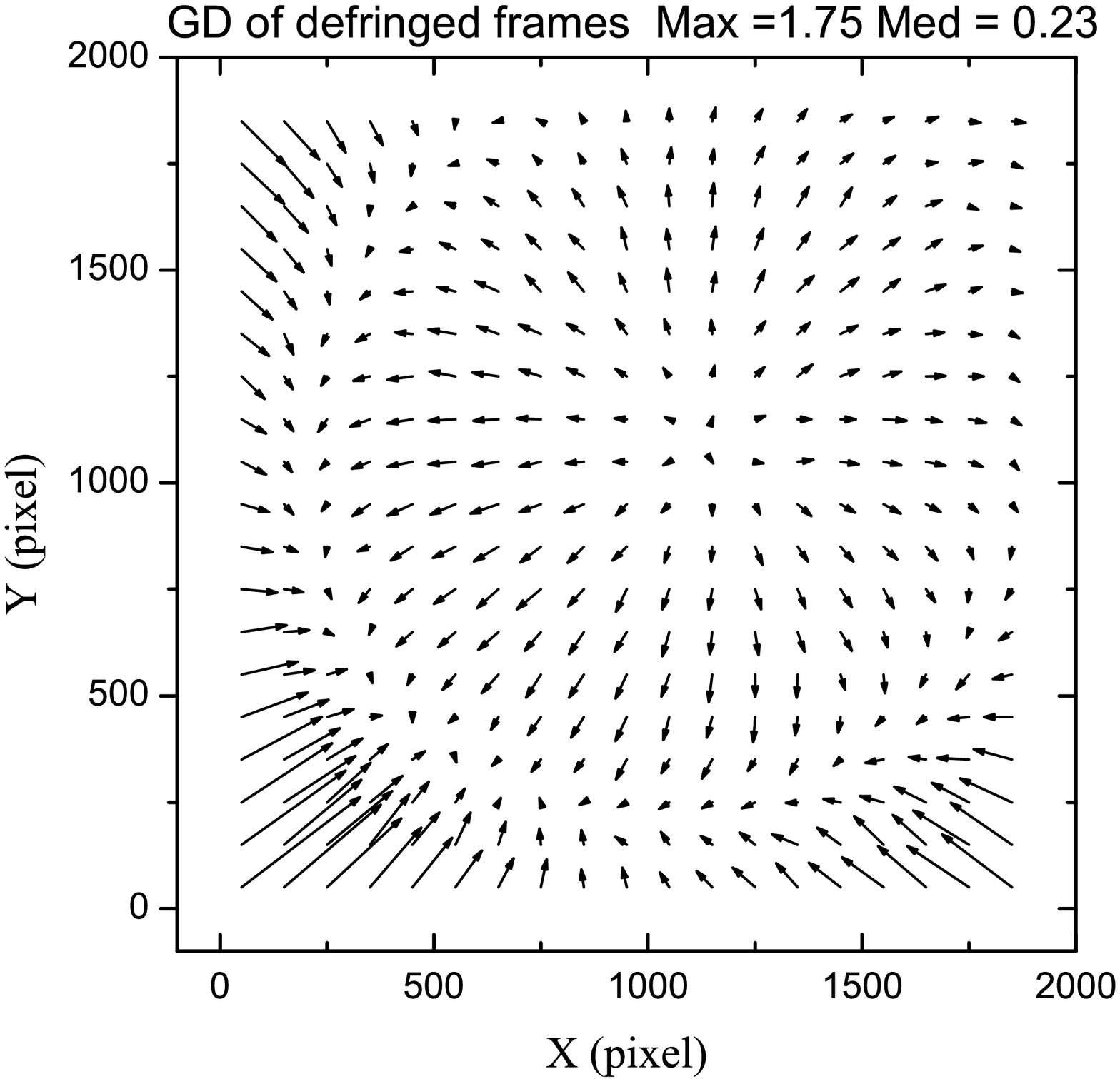}
\includegraphics[width=0.33\textwidth]{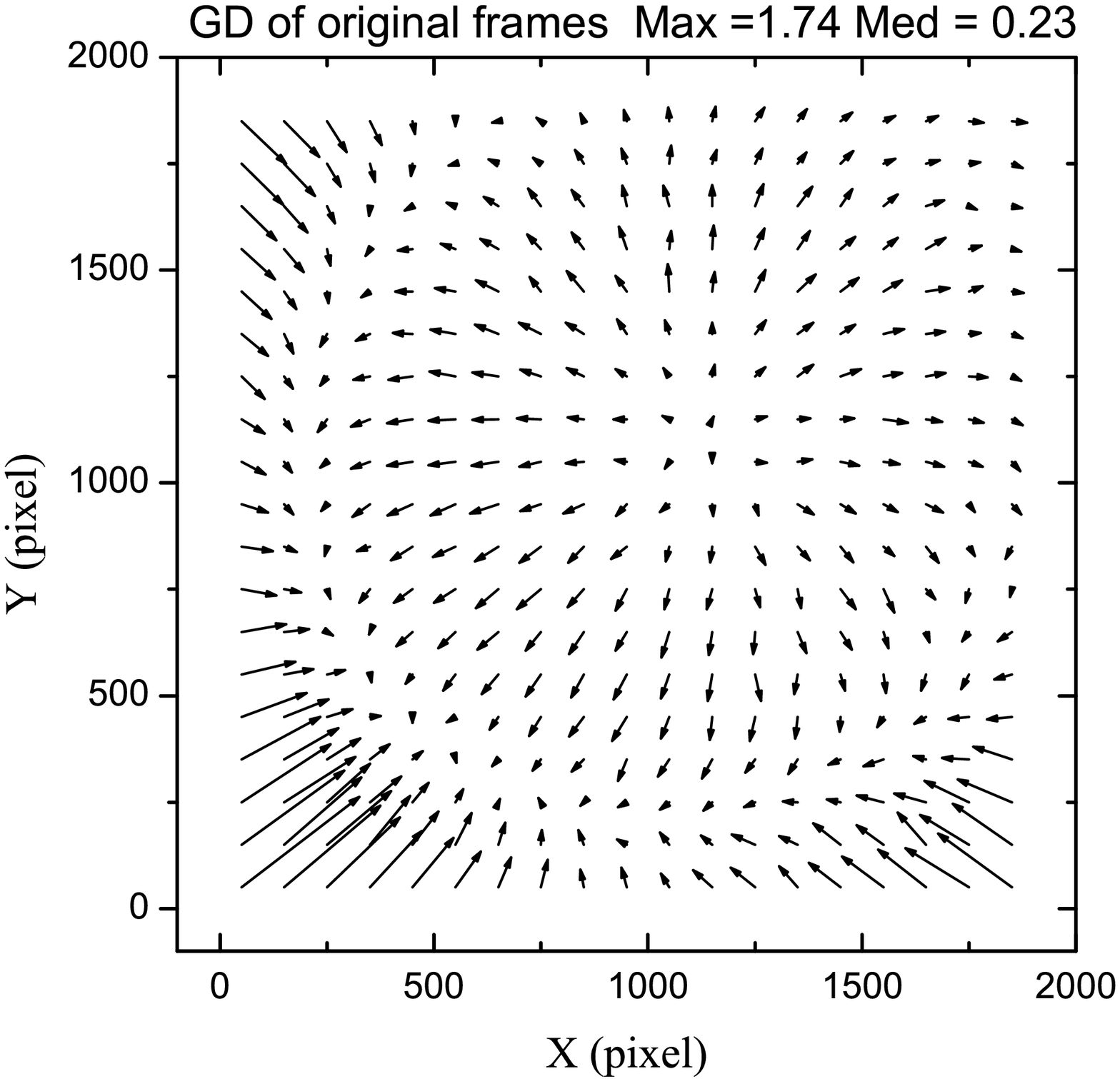}
\includegraphics[width=0.33\textwidth]{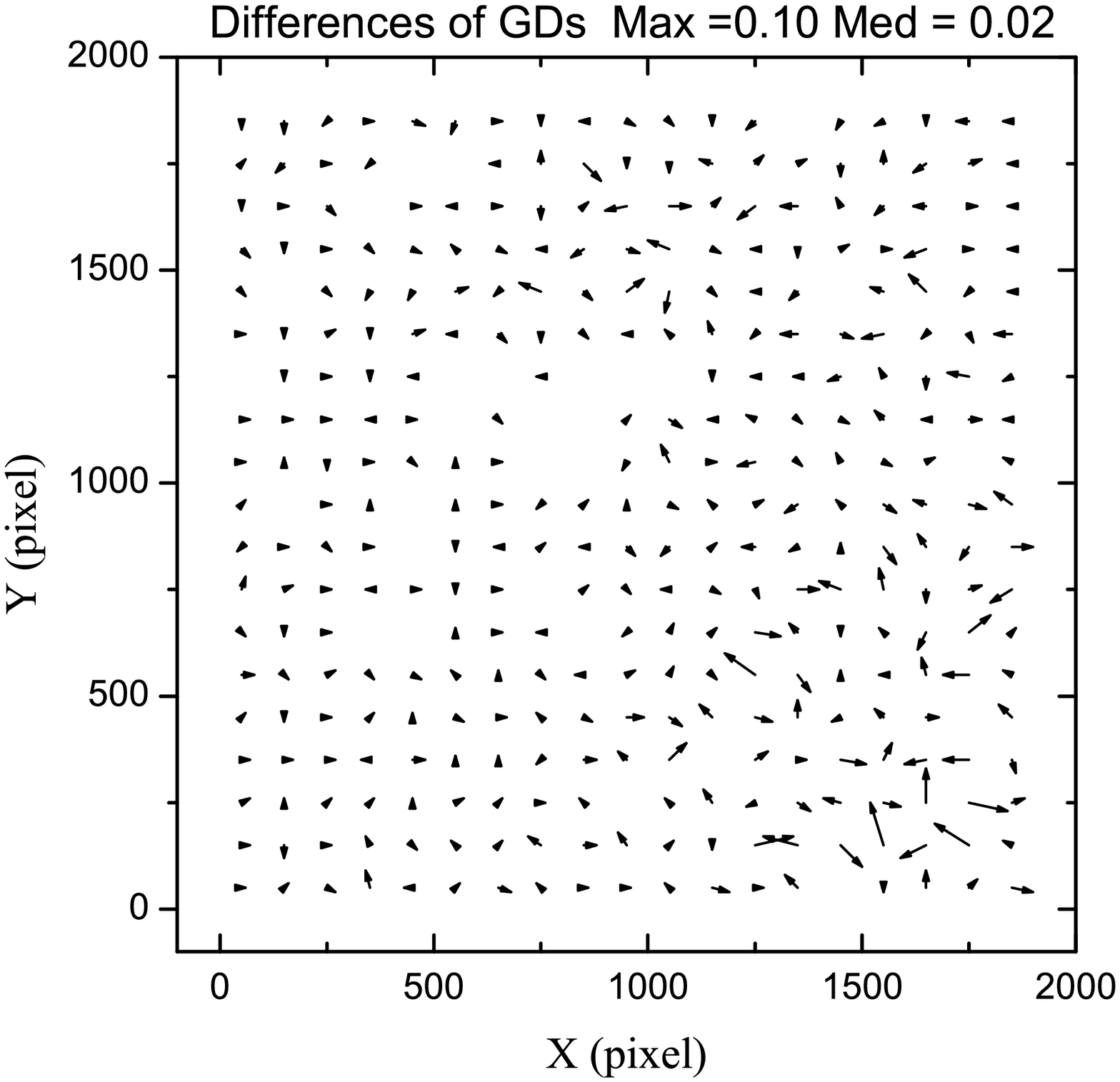}
\caption{The GD patterns are derived from the defringed frames and original frames respectively, as shown in the left and middle panel, magnified by a factor of 200. The right panel shows the differences of the two GD patterns, magnified by 1000. The units of Max and Med (Median) are in pixels.}
\label{Fig3}
\end{figure*}

\section{THE EFFECT ON ASTROMETRY}
\label{sect:astrometry}
We adopt the GD pattern derived from defringed frames to correct the pixel positions measured from the original and defringed CCD frames respectively. In order to study fringes' effects for faint stars' astrometric measurement, we need to analyze the results based on the magnitude $M_I$ of the PPMXL catalog (since the equivalent Johnson $I$ filter is used). It should be noted that, the source of $M_I$ is the $I$ magnitude from USNO-B and what's more, the USNO-B magnitude system is not recalibrated in PPMXL catalog as there are discrepancies in the magnitude system from field to field and from early to late epoch \citep{Roeser10}. Hence $M_I$ represents only a reference magnitude rather than a reliable standard magnitude for a star in our procedure. The magnitude $M_I$ is unknown for many faint stars in PPMXL catalog. According to the transformation of an observed instrumental magnitude to a standard system \citep{Da Costa92}, we assume there is a linear relationship between instrumental magnitudes and catalog magnitudes, which can be expressed as $M_I = a + b \times \overline{m}_{inst}$, where $\overline{m}_{inst}$ represents the average instrumental magnitude of a star measured more than once. We can solve $a$ and $b$ by a least-square fitting. The value of the slope $b$ is about 0.9 for both clusters and the intercept $a$ is about 0.3 for NGC 2324 and 1.1 for NGC 1664 respectively. The standard deviation (called SD hereafter) of the same stars' (O $-$ C) positional residuals for the two open clusters before and after defringing are shown in Figure~\ref{Fig4} and Figure~\ref{Fig5} as function of calculated $M_I$. We assume the faint stars would be more susceptible to the fringes and we roughly divide the bright and faint stars according to the difference before and after defringing in Figure 4 and 5. Similar divisions between the bright and faint stars are made in Section 5. Fringes reveal a greater effect for the faint stars ($M_I>15$). Moreover, the improvement is more significant at R.A. than at decl. We suppose it is mainly owe to fringes' clearly vertical trend (see the left panel of Figure~\ref{Fig2}}), and meanwhile the x axis is almost aligned with R.A. (a star-trailing operation is done before the observations). Based on the magnitude, we list the detailed statistics in Table~\ref{Tab2} and Table~\ref{Tab3}. For stars fainter than 15 in $M_I$, the improvement is significant in both directions. After defringing, the precisions in two coordinates are almost of the same level.

\begin{figure*}[t]
\centering
\includegraphics[width=0.4\textwidth]{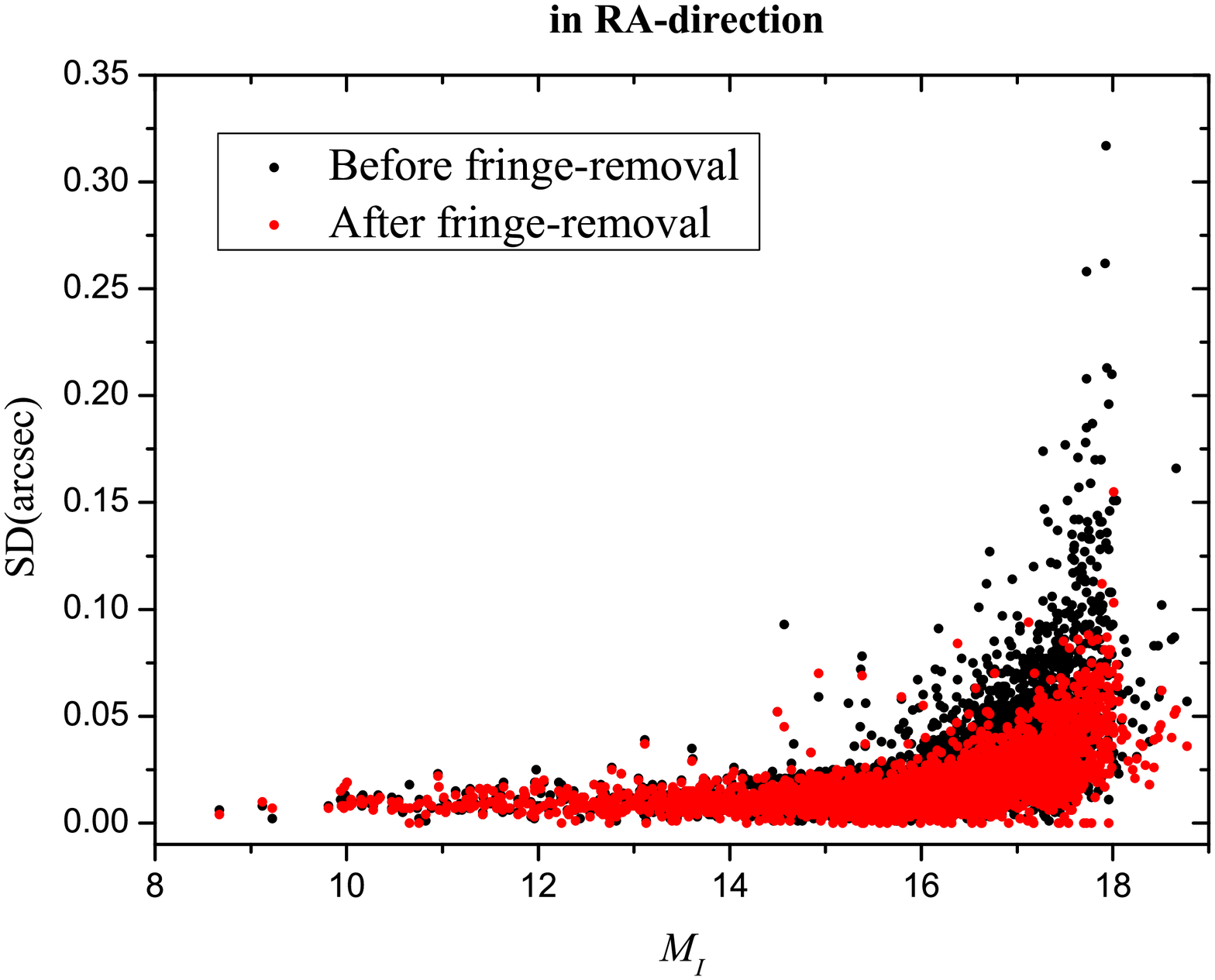}
\includegraphics[width=0.4\textwidth]{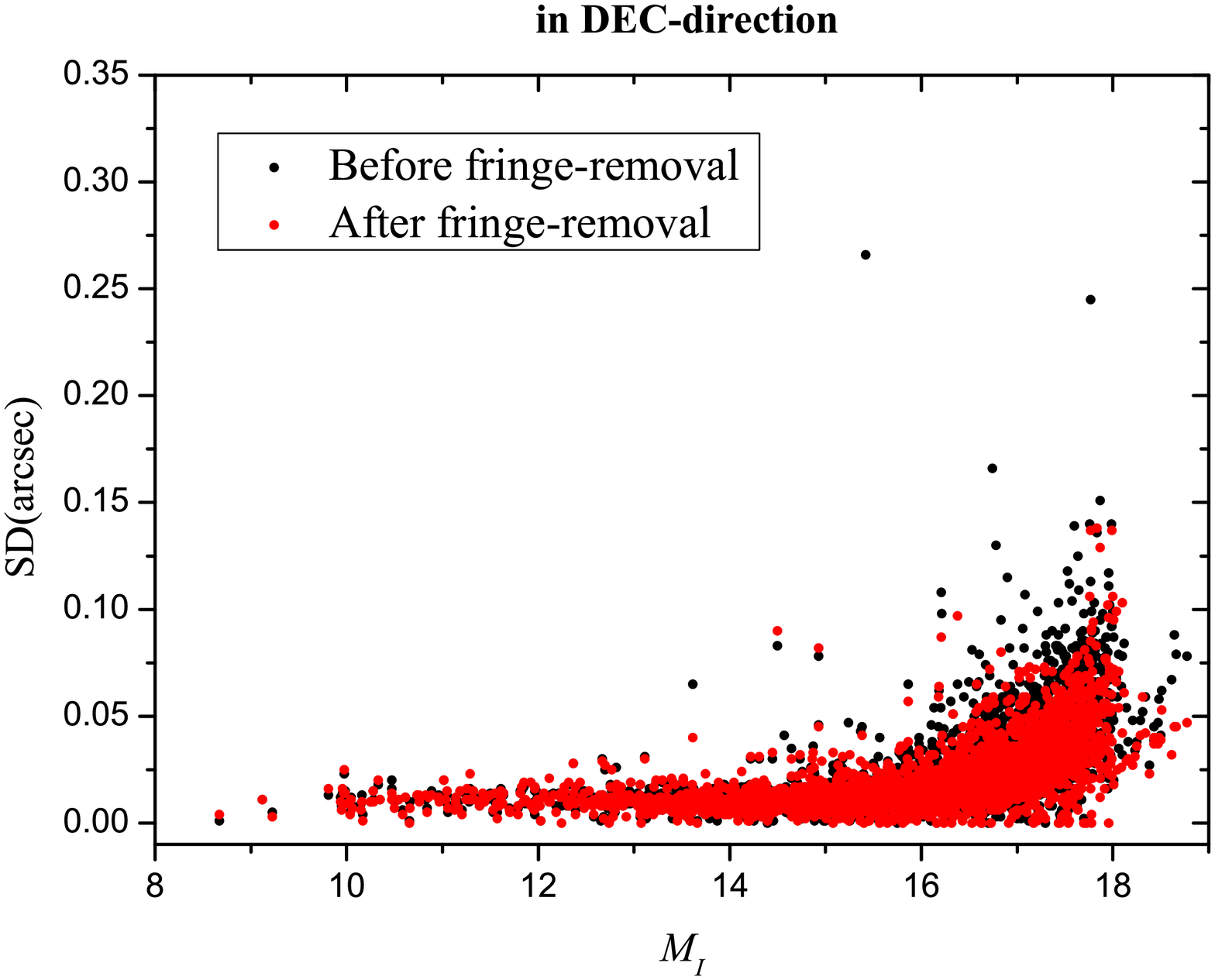}
\caption{SD of (O $-$ C) residuals for stars of NGC 2324 before and after defringing, with respect to $M_I$.}
\label{Fig4}
\end{figure*}

\begin{figure*} [t]\centering
\includegraphics[width=0.4\textwidth]{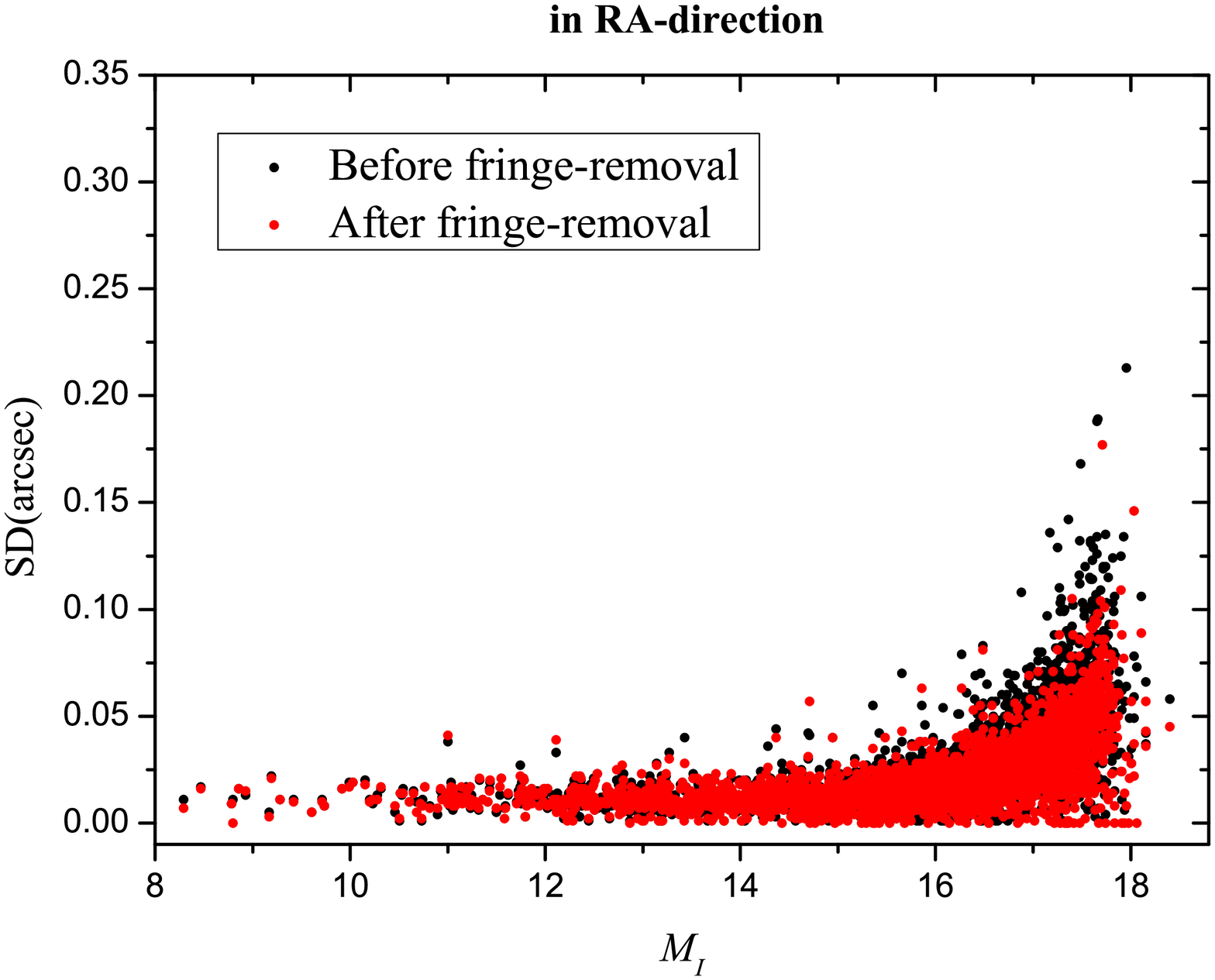}
\includegraphics[width=0.4\textwidth]{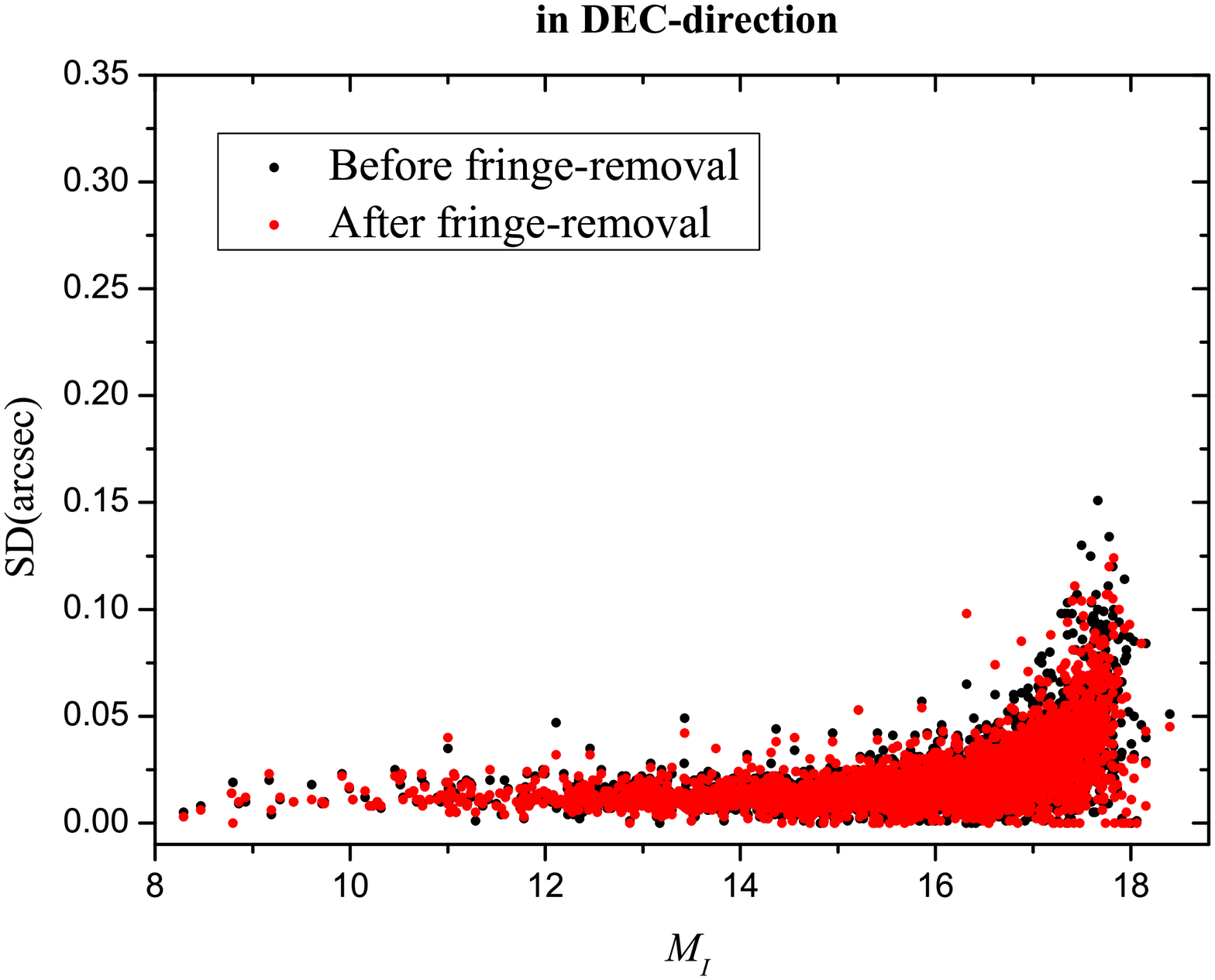}
\caption{SD of (O $-$ C) residuals for stars of NGC 1664 before and after defringing, with respect to $M_I$.}
\label{Fig5}
\end{figure*}
\begin{table}[t]
\begin{center}
\caption[]{Statistics of (O $-$ C) positional residuals of stars brighter than 15 in $M_I$ before and after defringing.  Column 1 presents the star cluster. Column 2 is the number of the stars in each cluster. The third column (`before' or  `after') shows the statistics before or after fringes' removal. The following two columns list the mean (O $-$ C) ($\mu_\alpha$) and its standard deviation ($\sigma_\alpha$) in right ascension. The last two columns list the mean (O $-$ C) ($\mu_\delta$) and its standard deviation ($\sigma_\delta$) in declination. All units are in milliarcseconds.}\label{Tab2}
%%Please Capitalize the First Letter of Each Notional Word in table's caption
 \begin{tabular}{ccccccc}
  \hline
cluster & N &  & $\mu_\alpha$ & $\sigma_\alpha$ & $\mu_\delta$ & $\sigma_\delta$ \\
  \hline\noalign{\smallskip}
\multirow{2}{*}{NGC 2324} & \multirow{2}{*}{773} & before &11	&6	&11	&6 \\
 & & after	&10	&5	&11	&6 \\
   \hline\noalign{\smallskip}
\multirow{2}{*}{NGC 1664} & \multirow{2}{*}{871} & before	&12	&5	&12	&5 \\
 & & after	&12	&5	&13	&5 \\
    \hline
\end{tabular}
\end{center}
\end{table}
%________________________________________ Table 3: Impacts of Astrometry for faint stars(I_J>15)
\begin{table}[t]
\begin{center}
\caption[]{Statistics of (O $-$ C) positional residuals of stars fainter than 15 in $M_I$ before and after defringing. The meaning of the columns are the same as Table~\ref{Tab2}.}\label{Tab3}
%%Please Capitalize the First Letter of Each Notional Word in table's caption
 \begin{tabular}{ccccccc}
  \hline
cluster & N &  & $\mu_\alpha$ & $\sigma_\alpha$ & $\mu_\delta$ & $\sigma_\delta$ \\
  \hline\noalign{\smallskip}
\multirow{2}{*}{NGC 2324} & \multirow{2}{*}{2953} & before &32	&27	&27	&20 \\
 & & after	&21	&14	&23	&15 \\
   \hline\noalign{\smallskip}
\multirow{2}{*}{NGC 1664} & \multirow{2}{*}{3021} & before  &30	&22	&25	&17 \\
 & & after	&25	&16	&24	&15 \\
    \hline
\end{tabular}
\end{center}
\end{table}

 \section{THE EFFECT ON PHOTOMETRY}
\label{sect:photometry}
We compare the magnitudes for the same star images measured on the original frames and the magnitudes measured on the defringed frames. Figure~\ref{Fig6} shows the differences (original minus defringed) change with the magnitudes (marked as ${m}_{inst}$). Detailed statistics are shown in Table~\ref{Tab4}. In a long exposure time as 30 seconds, a decrease of about 0.6\% in flux corresponds to a faint star image at ${m}_{inst}>18$. It shows that the fringes make an additive contribution to the flux counting of a star image as the fringes are formed by long-wavelength photon which is hardly absorbed by the antireflection coating of the CCD. And some bright star images' magnitudes have a clear difference after defringing. It is found that they locate in a crowed field with some faint neighbors. And we derive the photometric errors for the observations before and after defringing as following.
\begin{table}[t]
\begin{center}
\caption[]{Statistics of the differences in Figure~\ref{Fig6}. The results are derived for three groups depend on the magnitude of the star images, as shown in Column 2. Column 3 is the number of star images in each group. The following columns list the mean differences ($\mu$) and the standard deviation ($\sigma$) and the units are in magnitudes.}\label{Tab4}

%%Please Capitalize the First Letter of Each Notional Word in table's caption

 \begin{tabular}{ccccc}
  \hline
 cluster & ${m}_{inst}$ range & N & $\mu$ & $\sigma$ \\
  \hline\noalign{\smallskip}
\multirow{3}{*}{NGC 2324} & 10-16 & 9811 & 0.000 & 0.006\\
& 16-18 & 16537 & -0.001 & 0.022\\
& 18-20 & 5877 & -0.006 & 0.054\\
   \hline\noalign{\smallskip}
\multirow{3}{*}{NGC 1664} & 10-16 & 7882 & 0.000 & 0.004\\
& 16-18 & 15004 & -0.001 & 0.013\\
& 18-20 & 9300 & -0.001 & 0.035\\
    \hline
\end{tabular}
\end{center}
\end{table}

\begin{figure*} \centering
\includegraphics[width=0.4\textwidth]{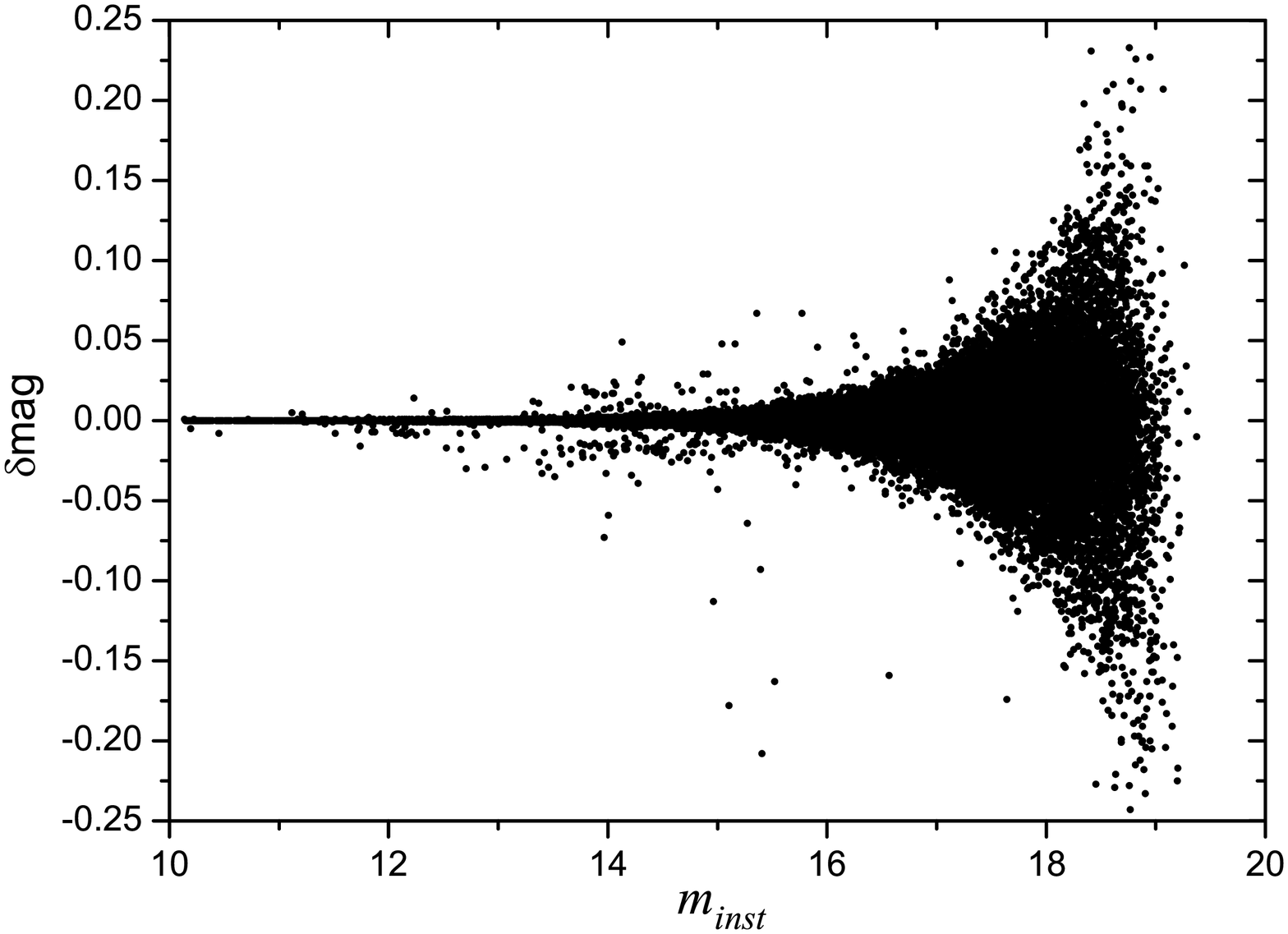}
\includegraphics[width=0.4\textwidth]{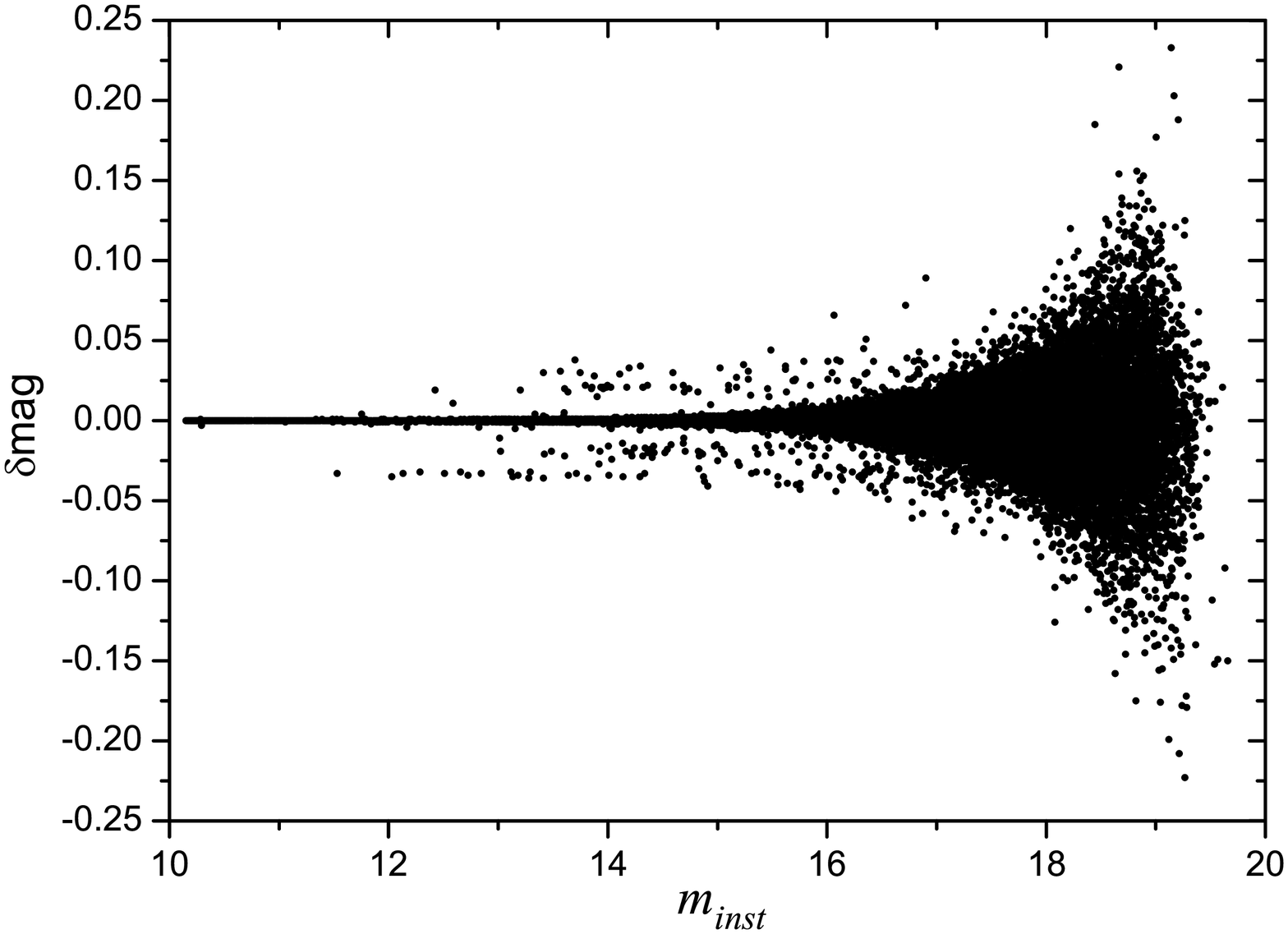}
\caption{Left: The difference in the instrumental magnitudes of the same star images before and after defringing for NGC 2324; right: the same for NGC 1664.}
\label{Fig6}
\end{figure*}
\begin{figure*} \centering
\includegraphics[width=0.4\textwidth]{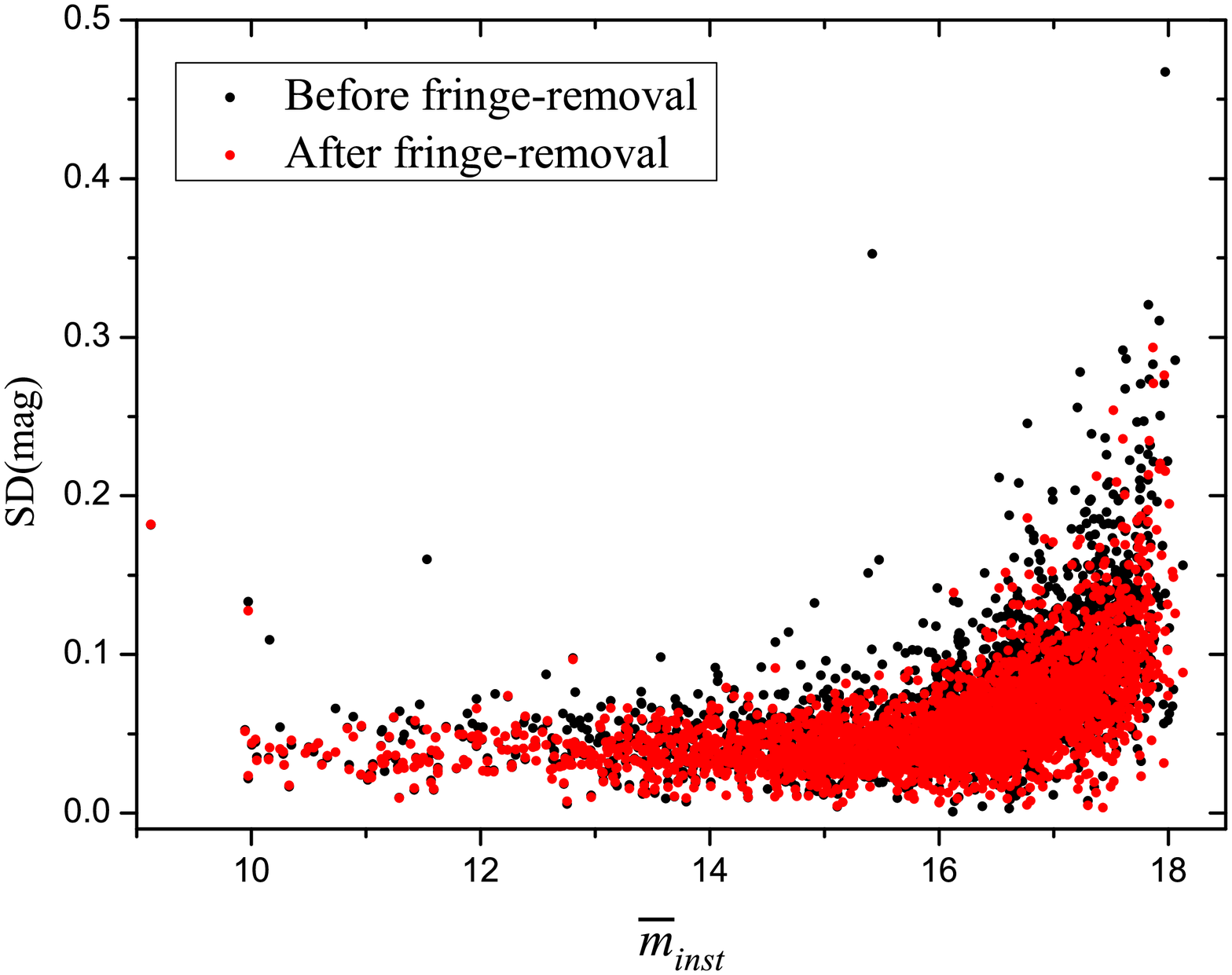}
\includegraphics[width=0.4\textwidth]{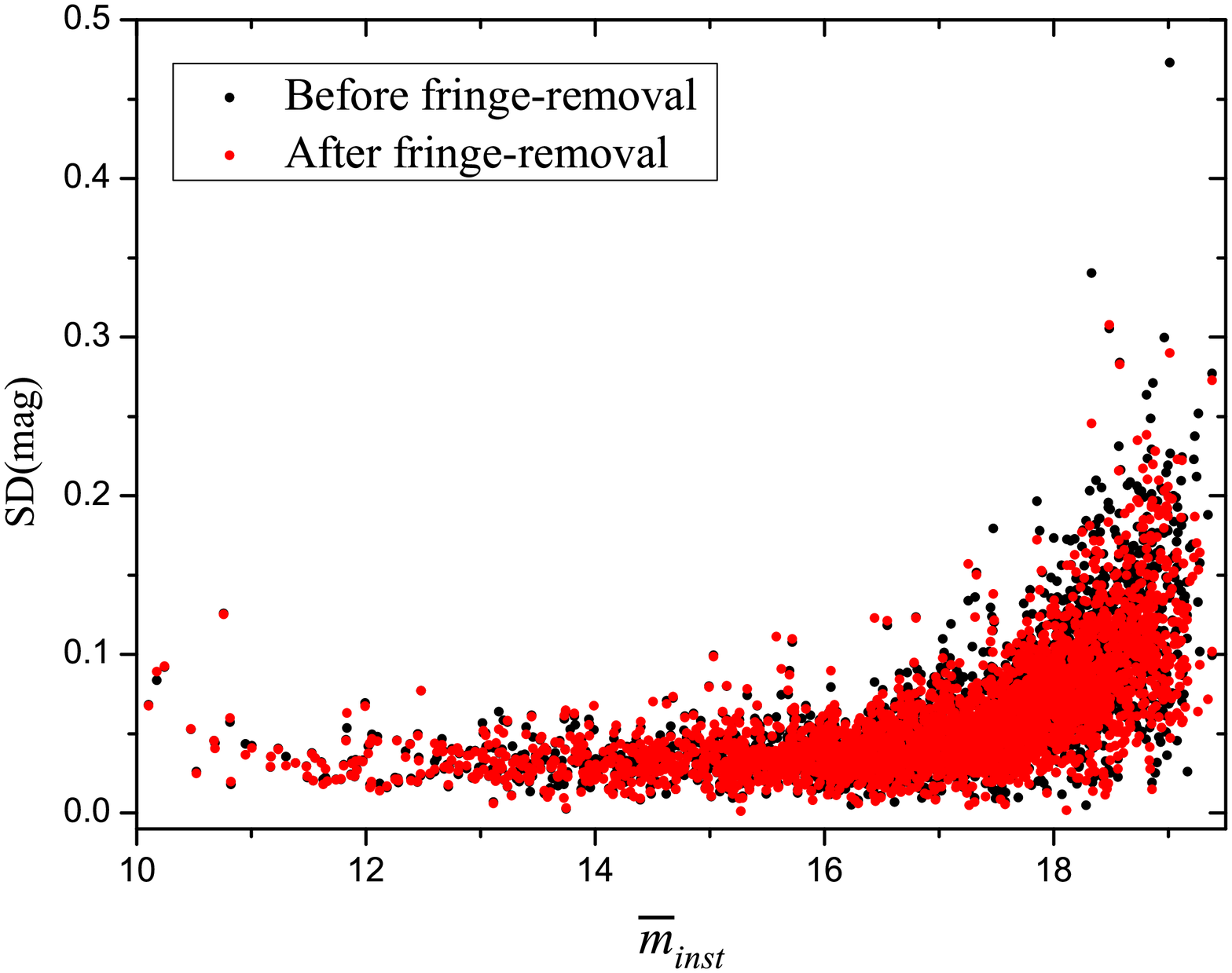}
\caption{Left: The photometric precision for NGC 2324; right: the same for NGC 1664.}
\label{Fig7}
\end{figure*}
Since the instrumental magnitude of the same star would vary with exposures for different observational environments, a standardization process is done as follows. Suppose there is a common star $k$ in the two neighboring exposures, $e_i$ and $e_j$. We compute the instrumental magnitude differences of many common stars in the two neighboring exposures and derive the mean value of these differences as shown in Equation~\ref{eq:2}. The mean of the instrumental magnitude difference is assumed as the baseline difference between the two exposures. For the star $k$, its instrumental magnitude in $e_j$ can be calibrated relative to $e_i$ and then to derive its photometric error in $e_j$ as Equation~\ref{eq:3}.
\begin{equation}
    \overline{\Delta}_{i,j} = \frac{1}{n}\sum\limits_{k=1}^n(mag_{i,k}-mag_{j,k})
  \label{eq:2}
\end{equation}
\begin{equation}
    err^{j}_{k} = mag_{j,k} + \overline{\Delta}_{i,j} - mag_{i,k}
  \label{eq:3}
\end{equation}
We calculate the standard deviation of the photometric errors for the same star in different frames and the results are shown in Figure~\ref{Fig7} as function of stars' average instrumental magnitude (noted as $\overline{m}_{inst}$). There is a more significant improvement for the photometric measurement of NGC 2324 after defringing since fringes appear more clearly with the increase of exposure time and make greater impacts on the background determination. The large discrepancy for some bright stars is due to saturation or blended stars. We analyze the standard deviation of the photometric errors dividually by stars' instrumental magnitudes as listed in Table~\ref{Tab5} and Table~\ref{Tab6}. The fringes' impacts on photometry are negligible for bright stars in short exposure time, while they grow up significantly for both bright stars and faint stars with long exposure time.
%________________________________________ Table 4: Impacts of Photometry for NGC 2324
\begin{table}[t]
\begin{center}
\caption[]{Statistics of photometric residuals for NGC 2324 before and after defringing. The results are derived for two groups as shown in Column 1 (`bright' or `faint'), which is composed of stars brighter or fainter than 16$^\mathrm{th}$ instrumental magnitude. Column 2 is the number of stars in each group. The third column (`before' or `after')  shows the statistics before or after fringes' removal. The following columns list the mean error ($\mu$) and its standard deviation ($\sigma$) for photometry and the units are in magnitudes.}\label{Tab5}

%%Please Capitalize the First Letter of Each Notional Word in table's caption

 \begin{tabular}{ccccc}
  \hline
 group & N &  & $\mu$ & $\sigma$ \\
  \hline\noalign{\smallskip}
\multirow{2}{*}{bright} & \multirow{2}{*}{1212} & before &0.046	&0.020 \\
 & & after	&0.041	&0.014 \\
   \hline\noalign{\smallskip}
\multirow{2}{*}{faint} & \multirow{2}{*}{1737} & before &0.083	&0.044 \\
 & & after	&0.070	&0.034 \\
    \hline
\end{tabular}
\end{center}
\end{table}
%________________________________________ Table 5: Impacts of Photometry for NGC 1664
\begin{table}[t]
\begin{center}
\caption[]{Statistics of photometric residuals for NGC 1664 before and after defringing. The results are derived for two groups as shown in Column 1 (`bright' or `faint'), which is composed of stars brighter or fainter than 17$^\mathrm{th}$ instrumental magnitude. The rest columns are the same as Table~\ref{Tab5}.}\label{Tab6}

%%Please Capitalize the First Letter of Each Notional Word in table's caption

 \begin{tabular}{ccccc}
  \hline
 group & N &  & $\mu$ & $\sigma$ \\
  \hline\noalign{\smallskip}
\multirow{2}{*}{bright} & \multirow{2}{*}{1207} & before &0.038	&0.016 \\
 & & after	&0.038	&0.016 \\
   \hline\noalign{\smallskip}
\multirow{2}{*}{faint} & \multirow{2}{*}{1731} & before &0.083	&0.044 \\
 & & after	&0.077	&0.039 \\
    \hline
\end{tabular}
\end{center}
\end{table}

\section{Conclusions}
\label{sect:conclusion}
We investigate the effects of fringes on astrometry and photometry based on 88 CCD frames of NGC 2324 and NGC 1664 which were taken at the 2.4-m telescope at Yunnan Observatory. After defringing, the astrometric precision of faint stars ($M_I>15$) has been significantly improved especially in R.A. direction, which is corresponding to x axis of the CCD plate, as the fringe pattern shows a greater trend in y direction. In a long exposure time as 30 seconds, a faint star image at ${m}_{inst}>18$ is about 0.006 magnitude fainter than before as the fringes make an additive contribution to the flux counting of a star image. Meanwhile, photometric errors are reduced by about 20\% for stars fainter than 16$^\mathrm{th}$ instrumental magnitude.

% Acknowledgements
\acknowledgments
We acknowledge the support of the staff of the 2.4-m telescope at Yunnan Observatory. Funding for the telescope has been provided by CAS and the People's Government of Yunnan Province. The work is financially supported by the National Natural Science Foundation of China (Grant No. U1431227, No. 11273014) and partly by the Fundamental Research Funds for the Central Universities. And our deepest gratitude goes to the anonymous reviewers for their careful reading and constructive suggestions that have helped improve this paper substantially.

%% References
%% Please cite all reference entries in the article text using \cite or
%% equivalent command.
%% Non-BibTeX  (Name-Year style)
%


\begin{thebibliography}{}
\bibitem[Da Costa(1992)]{Da Costa92} Da Costa, G. S. 1992, ASPC, 23, 90-104
\bibitem[Gullixson(1992)]{Gullixson92} Gullixson, C. A. 1992, \pasp, 23, 130
\bibitem[Howell(2012)]{Howell12} Howell, S. B. 2012, \pasp, 124(913), 263-267
\bibitem[Peng et al.(2012)]{Peng12} Peng, Q. Y., Vienne, A., \& Zhang, Q. F., et al. 2012, \aj, 144, 170
\bibitem[R\"oeser et al.(2010)]{Roeser10} R\"oeser, S., Demleitner, M., \& Schilbach, E. 2010, \aj, 139, 2440-2447
\bibitem[Snodgrass \& Carry(2013)]{Snodgrass13} Snodgrass, C., \& Carry, B. 2013, The Messenger, 152, 14
\bibitem[Wong(2010)]{Wong10} Wong, M. H. 2010, WFC3 ISR 2010-04
\bibitem[Zhang et al.(2012)]{Zhang12} Zhang, Q. F., Peng, Q. Y., \& Zhu, Z. 2012, Res. Astron. Astrophys., 12(10), 1451
 \end{thebibliography}
\end{document}